# A PHENOMENOLOGICAL APPROACH TO THE SIMULATION OF METABOLISM AND PROLIFERATION DYNAMICS OF LARGE TUMOR CELL POPULATIONS

*Roberto Chignola and #Edoardo Milotti

**Dipartimento Scientifico e Tecnologico, Universita' di Verona and Istituto Nazionale di Fisica Nucleare, Sezione di Trieste – Strada Le Grazie, 15 – CV1, I-37134 Verona, Italia*

*#Dipartimento di Fisica, Universita' di Udine and Istituto Nazionale di Fisica Nucleare, Sezione di Trieste – Via delle Scienze, 208 – I-33100 Udine, Italia*

**Abstract**

A major goal of modern computational biology is to simulate the collective behaviour of large cell populations starting from the intricate web of molecular interactions occurring at the microscopic level. In this paper we describe a simplified model of cell metabolism, growth and proliferation, suitable for inclusion in a multicell simulator, now **under development** (Chignola R and Milotti E 2004 *Physica A* **338** 261-6). Nutrients regulate the proliferation dynamics of tumor cells which adapt their behaviour to respond to changes in the biochemical composition of the environment. This modeling of nutrient metabolism and cell cycle at a mesoscopic scale level leads to a continuous flow of information between the two disparate spatiotemporal scales of molecular and cellular dynamics that can be simulated with modern computers and tested experimentally.

## 1. Introduction

Biological systems span vast spatiotemporal scales, from the microscopic dynamics of atoms to the macroscopic dynamics of cell clusters. Information flows in both directions and determines the behaviour of living matter and ultimately the normal physiology of organisms and the onset of pathologies such as tumors.

The malignancy and the response of tumors to therapy is dependent upon their growth potential which in turn is determined by the ability of tumor cells to adapt to different tissue environments, to compete with normal cells for both space and nutrients and to ignore molecular signals attempting to block cell cycling or to promote cell death [1]. Major efforts have been done by experimenters to highlight the molecular circuits underlying tumor cell biology and, on this basis, to develop novel therapeutic strategies. This has resulted in a huge body of knowledge which has deepened our understanding of the molecular details of tumor cell biology but often with little or no consequence on the clinical management of tumors. Clinicians, in fact, deal with the macroscopic properties of tumors, i.e. masses as large as grams and that may grow for months or years, and hence with spatiotemporal scales which in general are not explored in current molecular experimental setups. Computer simulations might allow to fill this gap, but they are very hard to setup and run, and an ongoing challenge in the field of biological computer simulation involves the treatment of disparate time and length scales [2]. Current methods are limited as even the most powerful techniques from molecular dynamics allow one to simulate only very small objects (of the order of $10^5$ atoms) for very short time spans (of the order of nanoseconds), much shorter than it would be required for the simulation of tumors of clinical interest [3].

We have proposed and started the development of a numerical simulator of tumor growth and we have already achieved a linearization of time and memory requirements, using techniques and methods borrowed from molecular dynamics and computational geometry [3]. **The simulator should eventually reproduce the growth of solid tumors in the prevascular phase and allow an *in silico* investigation of the biophysical laws governing tumor growth dynamics and the response of cell clusters to anti-tumor treatments. The program simulates both cells and the complex and changing environment where molecules, such as nutrients and drugs, diffuse and where cells interacts with each other through viscoelastic forces [3]. This involves different spatiotemporal scales where information flows and connects the biochemical fluctuations of the environment to the variations in the growth of a multicell population. The simulation requires a subprogram to describe cell metabolism and cell cycle, which links the information carried by molecules in the environment to cell metabolism and vice versa.**

**In this paper we describe the simulator of cell metabolism and of cell cycle dynamics that shall be incorporated into a full simulator of tumor spheroid growth. The simulator takes into account the topology of the cell** metabolic network, with some of the metabolic steps actually known and measured, which is then integrated into a model of the cell cycle. It is nearly impossible to model the whole molecular network underlying cell metabolism and the control of the cell cycle for large cell populations, since this would mean: a) specializing the model to a specific cell line (or even to a single cell if one includes individual and environmental variability); b) having to deal with an incomplete model, since the network connections and parameters are not completely known for most cell lines; c) increasing the complexity of the cell model to

a level that might lead to an excessive computational load. **Thus, certain molecular pathways whose fine details are not directly relevant to cell behaviour and to cell-environment interactions have been modelled phenomenologically.**

On the other hand a reduced network structure is justified by: a) network robustness, i.e., the metabolic network is dominated by node hierarchy as described in [4] and is not sensitive to the actual topology, but only to the dominance of certain network nodes (hubs) that we hope to have correctly identified; b) function interchangeability and redundancy, i.e., some metabolic paths perform similar functions by means of different enzymes or groups of enzymes; c) parameter adjustment, i.e., the procedure by which the parameters of the simulated network are adjusted to fit the macroscopic cell behavior: this means that some parameters may be different from the actual measured values, but accomodate for missing branches in the metabolic network; d) statistical insensitivity in large cell populations. The last point deserves further explanation: from the experience with large statistical ensembles in Thermodynamics and Statistical Mechanics, we know that many fine details of a complex physical system may be skipped without sacrificing precision in the description of its macroscopic behavior. For example the ideal gas, with the inclusion of the Van der Waals terms, is an exceptionally good description of the behavior of real gases in many different physical conditions, and a couple of parameters suffice to parametrize the vast variety of atomic and molecular gases and vapors, and here we aim at a reduced network structure that behaves in a similar fashion.

We show that our reduced model of cell metabolism and of the cell cycle can simulate the growth dynamics of tumor cell populations with good quantitative agreement between simulated and experimental variables. Since the model uses real parameter values, **either estimated from experiments or fitted to experimental data**, it can be exploited to develop *in silico* studies of **tumor growth** to support experimental research.

## 2. Experimental Section

*2.1 Simulations*

The model described in the present paper has been written in **ANSI-C** and has been run on an Apple PowerMac G4 and on an Apple Powerbook G4, under the operating system OS X (the Apple flavour of the UNIX operating system).

In all cases the difference equations have been solved iteratively with $\boldsymbol{\Delta t}$ = 1 s; this choice is dictated by the opposite needs of reducing the total computation time as much as possible and by the computational request – needed for algorithmic stability – that $\Delta t \ll \tau_k$ for all the Michaelis-Menten (MM) reactions, where $\tau_k$ is the reaction time constant for low concentrations, for the $k$-th MM reaction. Where needed, pseudorandom numbers have been generated by the linear congruent generator RAN2, taken from the Numerical Recipes library [5].

*2.2 Cell lines and cytofluorimetric assays*

Molt3, CaCo2 and Whei164 cells were obtained by the ATCC and maintained in RPMI-1640 medium supplemented with 10% foetal bovine serum. Cells were cultured at 37°C in a 5% $CO_2$ humidified atmosphere and passaged weekly. Human peripheral mononuclear cells were drawn from healthy volunteers as described previously [6]. Cells from exponentially growing cultures were labelled with ethidium bromide and analysed as described [6] using an EPICS-X (Coulter, Hialeah, FL) cytofluorimeter. Human T lymphocytes were stimulated to proliferate by adding 1% phytohemagglutinin (PHA, Sigma) to culture wells. Proliferation was monitored in parallel by $^3$H-thymidine incorporation assays [6].

## 3. Modelling tumor cell metabolism

Both normal and tumor cells use different fuels to produce energy, such as sugars, aminoacids and lipids. One of the best studied is certainly glucose. There are many experimental data on glucose deprivation from growth media and its effect on cell proliferation, and they may be used to test the results of numerical simulations. In addition, tumor cells are in general believed to depend on glucose catabolism more than normal ones [7-9]. This difference in metabolic activity between normal and cancerous cells is now being exploited and used to diagnose and monitor cancer and to treat it [10,11]. Since we are interested in the simulation of tumor cell clusters [3] we concentrate here on modeling glucose uptake and utilization by these cells. However we also consider other metabolic fuels since a comprehensive, though minimal, simulator of cell metabolism cannot be based solely on glucose. The metabolic network that we have implemented is shown in Fig.1; each biochemical step will be explained in the following paragraphs, while the model parameter values are listed in Tab.1.

### *3.1 Glucose uptake and retention by cells*

Glucose is actively transported across cell membranes by means of a family of transporter molecules expressed at the cell surface collectively called GLUT [12]. The major member of the GLUT family expressed by tumor tissues is GLUT1 and other members appear to follow a tissue-specific expression pattern [12,13]. The GLUT receptors can work in both directions by pumping glucose inside and outside the cell. The rates of these processes (rates $v_{1p}$ and $v_{1m}$ in Fig.1) are saturable as a function of glucose concentration and their kinetics have been modelled using Michaelis-Menten-like equations whose general form is:

$$\frac{dC}{dt} = v \qquad (1a)$$

$$v = \frac{V_{max} \cdot C}{K_m + C} \qquad (1b)$$

where, as usual, $C$ is the concentration of the substrate molecules, $V_{max}$ is the maximum rate of the process (which in this case depends on GLUT transporters concentration) and $K_m$ is a constant that corresponds to the concentration of $C$ at which the process has rate $V_{max}/2$.

Under hypoxic conditions the uptake of glucose by cells is increased up to 2.3 fold [14,15]. It has been recently shown that the expression of the GLUT1 gene is upregulated under the control of the hypoxia-inducible transcription factor HIF-1 [16]. Also, it is known that glucose uptake is increased in hypoxic tumor tissues [13]. We conclude that the overexpression of GLUT transporters at the cell surface should increase the maximal rate of glucose transport, thus, in modelling the uptake of glucose by cells we have introduced a parameter $h$, which tunes the $V_{max}$ as a function of the intracellular oxygen concentration:

$$\frac{dG_{ex}}{dt} = v_{1m} - v_{1p} \qquad (2a)$$

$$v_{1p} = \frac{V_{max1} \cdot G_{ex}}{K_{m1} + G_{ex}} \qquad (2b)$$

$$v_{1m} = \frac{V_{max1} \cdot G_{in}}{K_{m1} + G_{in}} \qquad (2c)$$

$$V_{max1} = V_m \cdot h \cdot Surf \qquad (2d)$$

$$h = \begin{cases} 1 & \text{if } ConcO_2 > O_2st \\ 1.3 \cdot (1 - \frac{ConcO_2}{O_2st}) + 1 & \text{elsewhere} \end{cases} \qquad (2e)$$

where $G_{ex}$ and $G_{in}$ are the external and the intracellular glucose concentrations, respectively, $V_m$ and $K_{m1}$ are the Michaelis-Menten constants of glucose transport by GLUT1 receptors, $Surf$ is the cell surface area, $ConcO_2$ is the intracellular oxygen concentration and $O_2st$ is the standard atmospheric oxygen concentration (see below).

Once taken up from the environment, glucose is rapidly phosphorylated to form glucose-6-phosphate (G6P). Since G6P cannot bind the GLUT transporters, the phosphorylation prevents its active transport outside the cell. Two enzymes are known to phosphorylate glucose: glucokinase, which is mainly expressed by liver and pancreatic tissues and hexokinase which is expressed by non-liver tissues. The two enzymes differ in their enzymatic activities as the Km of glucokinase is much higher than that of exokinase [17]. **Thus, the activity of the two enzymes depends on the concentration of intracellular glucose.** Tumor cells have been found to express both enzymes [18]; **in addition, a gradient of nutrients is observed within tumor cell clusters and cells are therefore exposed to different glucose concentrations depending on their localization into the tumor mass [3,14] (see also references cited therein).** Both enzymatic processes have therefore been considered here (rates $v_2$ and $v_{22}$ for glucokinase and hexokinase, respectively, in Fig.1). The time-dependent variations of the intracellular glucose concentration are:

$$\frac{dG_{in}}{dt} = v_{1p} - (v_{1m} + v_2 + v_{22}) \qquad (3a)$$

$$v_{1p} = \frac{V_{\max 1} \cdot G_{ex}}{K_{m1} + G_{ex}} \qquad (3b)$$

$$v_{1m} = \frac{V_{\max 1} \cdot G_{in}}{K_{m1} + G_{in}} \qquad (3c)$$

$$v_2 = \frac{V_{\max 2} \cdot G_{in}}{K_{m2} + G_{in}} \qquad (3d)$$

$$v_{22} = \frac{V_{\max 22} \cdot G_{in}}{K_{m22} + G_{in}} \qquad (3e)$$

Both hexokinase and glucokinase expression and activity are downmodulated by low intracellular concentrations of glucose [19]. Downmodulation of enzyme activity has been realized by considering a Michaelis-Menten-like process relating the maximal rates of *G6P* formation ($V_{\max2}$ and $V_{\max22}$ in equations 3) to the intracellular concentration of glucose $G_{in}$:

$$V_{\max 2} = V_{m2} \cdot \frac{G_{in}}{K_a + G_{in}} \qquad (4a)$$

$$V_{\max 22} = V_{m22} \cdot \frac{G_{in}}{K_a + G_{in}} \qquad (4b)$$

where $V_{m2}$ and $V_{m22}$ are the experimentally-determined maximal rates of glucokinase and hexokinase activity, respectively, and $K_a$ is a parameter that assumes positive values.

The sets of equations (2)-(4) model the cellular uptake of glucose from the surrounding environment and its retention within cells as G6P.

### *3.2 Glucose-6-Phosphate utilization by cells and the metabolism of nutrients other than glucose*

There is a huge body of literature on G6P utilization and conversion into glycogen storage molecules (STORE compartment) from which we have sorted out a few simple relationships (Fig.1). Experimental data indicate that these relationships actually exist if we consider the cell as a whole, and not its detailed metabolic mechanisms [8]. It appears that a cell consumes a fixed fraction of glucose and/or glycogen molecules to produce energy and this fraction seems to be quite constant even when cells are cultured in media containing different glucose concentrations [8]. We wish to stress that these observations have not been directly reported in the cited papers, but rather they have been inferred from published data, and must be validated by comparing the outputs of simulations with further experimental observations.

The rates $g_1$, $g_2$ and $g_3$ of G6P conversion and the rates $r_1$, $r_2$ and $r_3$ of the glycogen stock consumption have therefore been modeled, in their general form, as follows:

$$r_i, g_i = q_i \cdot X_i \qquad (5)$$

where $X_i$ are either *G6P* or *STORE* molecules and $q_i$ are the fractions of these molecules utilized per unit time (these values must be tuned individually to match experimental data).

As shown in Fig.1, the rates $g_2$, $r_2$ and $r_3$ parametrize the velocity of the conversion of *G6P* and of *STORE* molecules into energetic ATP ones through oxidative phosphorylation. These rates, therefore, must vary as a function of the intracellular oxygen concentration. The rates $g_2$ and $r_2$ have been assumed to describe ATP production (rate *ATP_OX* in Fig.1) under normal environmental conditions, i.e. at standard extracellular glucose and oxygen concentrations. If the external glucose concentration decreases, the synthesis of G6P and the production rate *ATP_OX* decrease as well. Cells then react to low glucose concentrations by metabolizing more of the stored glycogen molecules [20], a process that has been modeled by means of the rate $r_3$.

In these cases, the STORE variable is progressively reduced and this reservoir of energetic molecules may eventually be exhausted.

To prevent complete consumption of the STORE compartment we have taken into account the gluconeogenesis pathway. Gluconeogenesis is an energy-dependent process and therefore requires the presence of ATP [17]. Molecules other than glucose can also be directly catabolized through the oxidative phosphorylation pathways to produce energy. Since numerous but different molecules - each one involving different enzymatic mechanisms - can participate in these processes we have introduced a generic compartment denoted by *A*. The *A* compartment, consists by definition of non-glucose molecules **(e.g. lactate, glutamine and other aminoacids)** that diffuse into cells and that can **contribute** to the oxidative phosphorylation and/or **contribute** to glycogen formation following the gluconeogenesis pathway (see the processes identified by rates $p_{11}$ and $p_{22}$ in Fig.1).

In the model, cells choose between gluconeogenesis and oxidative phosphorylation of molecules in *A* on the basis of the ATP production rate, which depends in turn on oxygen availability, **and hence the utilization and/or consumption of metabolites *A* is simplified through the switch portraied in Fig.1**. **This part of the simulation program will require further refinement to take into account e.g. processes of re-utilization of produced lactate upon its uptake by cells from the surrounding environment.**

Overall, these considerations lead one to take into account the existence of both oxygen and ATP sensors which continuously monitor the variations of the nutrient composition in the surrounding environment. Thus, before describing the equations that model variables *G6P*, *STORE* and *A*, a description of both sensors must be detailed.

### *3.3 Modelling the ATP and oxygen sensors*

The ATP sensor has been modelled by considering a threshold in the rate of ATP production by oxidative phosphorylation (*ATP_St*). The threshold has been fixed at values right below the rate of *ATP_OX* production under normal environmental conditions which is given by the following equation:

$$ATP_OX = 30 \cdot (g_2 + r_2) \cdot (\frac{MW_{ATP}}{MW_G}) \qquad (6)$$

where $MW_{ATP}$ and $MW_G$ are respectively the molecular weights of ATP and glucose, and the multiplicative factor 30 takes into account the stoichiometry of ATP production from 1 mole of glucose through the oxidative phosphorylation if we take a ratio P/O=2.5 [8].

With these definitions the ATP sensor is given by the following formulas:

$$\begin{cases} r_3 = 0 \\ p_{11} = f(r_1, d) \\ p_{22} = 0 \end{cases} \qquad if\, ATP_OX > ATP_St$$

$$(7)$$

$$\begin{cases} r_3 = c \cdot SensO_2 \cdot (ATP_St - ATP_OX) \cdot \dfrac{MW_G}{MW_{ATP}} \cdot \dfrac{1}{30} \\ p_{22} = d \cdot SensO_2 \cdot (ATP_St - ATP_OX) \cdot \dfrac{MW_G}{MW_{ATP}} \cdot \dfrac{1}{30} \\ p_{11} = f(r_1, d) - p_{22} \end{cases} \qquad elsewhere$$

where $f(r_1,d)$ and $c$ are functions that will be described in the next paragraph and $SensO_2$ is the output of the oxygen sensor that is explained below. It follows from equations (7) that if the rate of ATP production through oxidative phosphorylation of glucose and glycogen is above the threshold *ATP_St*, then no more glycogen molecules are metabolized and non-glucose molecules are degraded to balance the loss of glycogen from the STORE compartment. As a consequence the STORE compartment may reach a steady state. On the contrary, if the ATP production rate is below the threshold

*ATP_St* then: 1) additional STORE molecules are catabolized with kinetics proportional to the rate of ATP loss (rate *ATP2* in Fig.1) and, 2) non-glucose molecules are also metabolized to produce ATP (rate *ATP3* in Fig.1). Since the rate $p_{11}$ decreases proportionally to $p_{22}$ the STORE compartment receives less molecules through the gluconeogenesis pathway and shrinks, while the cell may transiently overshoot ATP production (i.e. *ATP_OX+ATP2+ATP3>>ATP_St*) and hence oxygen consumption. This part of the model has been designed to simulate experimental data showing higher oxygen consumption and ATP production by cells growing in low glucose-containing media [21,22].

ATP production via oxidative phosphorylation and oxygen consumption are correlated. The overall rate of oxygen consumption is:

$$k_0 = 6 \cdot (g_2 + r_2 + r_3 + p_{22}) \cdot \frac{MW_{O2}}{MW_G} \qquad (8)$$

where the multiplicative factor 6 takes into account the stoichiometry of oxygen consumption from 1 mole of glucose. If the amount of available oxygen molecules in the cell volume is $nO_2$, then:

$$k_d = nO_2 - k_0 \cdot dt \qquad (9)$$

measures at each time step the difference between the amount of available oxygen molecules and those consumed during ATP production. These calculations provide the basis to model the oxygen sensor:

$$\begin{cases} SensO_2 = 1 \;\; if \;\; k_d \geq 0 \\ SensO_2 = \dfrac{nO_2}{k_0 \cdot dt} \;\; elsewhere \end{cases} \qquad (10)$$

It should be noted that in the present model the equilibrium between the external and the intracellular oxygen concentrations is assumed at all time steps and that no diffusive processes have been considered. This is reasonable since the diffusion of oxygen through cellular membranes is much faster than the process of oxygen consumption (see the appendix and ref.[23]).

*3.4 Modelling the variations in the cellular content of G6P, STOCK and non-glucose A molecules*

From the considerations above and equations (3) and (4), the complete set of equations that describe the *G6P* time evolution can be written as follows:

$$\frac{dG6P}{dt} = v_2 + v_{22} - g_1 - g_2 - g_3 \qquad (11a)$$

$$g_1 = coeffg_1 \cdot G6P \qquad (11b)$$

$$g_2 = coeffg_2 \cdot SensO_2 \cdot G6P \qquad (11c)$$

$$g_3 = coeffg_3 \cdot G6P \qquad (11d)$$

where the coefficients $coeffg_i$ are listed in Tab.1.

The equations that model the variations of the *STORE* variable are:

$$\frac{dSTORE}{dt} = g_3 + p_{11} - (r_1 + r_2 + r_3) \qquad (12a)$$

$$r_1 = coeffr_1 \cdot c \qquad (12b)$$

$$r_2 = g3 \cdot c \cdot SensO_2 \qquad (12c)$$

where the rate $r_3$ is the same as in equations (7) and $coeffr_1$ is a parameter (see supplemetnal material).

The function *c* describes a homeostatic loop. As discussed above, *STORE* can eventually drop to zero depending on glucose and oxygen concentrations. We have assumed that in this case the rates $r_1$, $r_2$ and $r_3$ should progressively decrease as well, with a lower limit equal to zero. Thus, we defined the function *c* as follows:

$$c = \frac{STORE}{K_c + STORE} \qquad (13)$$

A similar homeostatic control loop has been introduced for the non-glucose nutrients in compartment *A* whose variations are described by the following equation:

$$\frac{dA}{dt} = -(p11 + p22) \qquad (14)$$

where $p_{11}$ and $p_{22}$ are the same as in equations (7). The function $f(r_1,d)$ in equations (7) incorporates the homeostatic control function *d* and is defined as follows:

$$f(r_1, d) = coeffp_{11} \cdot r_1 \cdot d \qquad (15a)$$

$$d = \frac{A}{[K_d \cdot Volume(t)] + A} \qquad (15b)$$

where $coeffp_{11}$ and $K_d$ are positive parameters whose values are given in Tab.1.

*3.5 ATP production and consumption*

Cells produce ATP through the anaerobic glycolysis and through the complete oxidation of nutrients. These processes are described (see Fig.1) by rates *ATP_nOx*, for anaerobic glycolysis, and *ATP_Ox*, *ATP2* and *ATP3* for the oxidative phosphorylation of glucose and of other nutrients. Recalling that 2 moles of ATP are produced per mole of glucose during anaerobic glycolysis and that 30 moles of ATP are produced per mole of glucose, or of glucose-equivalents, during complete oxidation of this substrate, the complete set of equations describing the rates of ATP production is:

$$ATP_nOx = 2 \cdot (g_1 + r_1) \cdot \frac{MW_{ATP}}{MW_G} \qquad (16a)$$

$$ATP_Ox = 30 \cdot (g_2 + r_2) \cdot \frac{MW_{ATP}}{MW_G} \qquad (16b)$$

$$ATP2 = 30 \cdot r_3 \cdot \frac{MW_{ATP}}{MW_G} \qquad (16c)$$

$$ATP3 = 28 \cdot p_{22} \cdot \frac{MW_{ATP}}{MW_G} \qquad (16d)$$

where $MW_{\mathrm{ATP}}$ and $MW_{\mathrm{G}}$ are the molecular weights of ATP and of glucose, respectively. The first equation also matches the rate of lactate production which is, if we take a ratio of 2 moles of lactate per moles of glucose:

$$AcL = 2 \cdot (g1 + r1) \qquad (17)$$

Equation (16d) takes into account **that molecules other than glucose (i.e. metabolites *A*) do not enter the glycolysis and hence they produce less ATP upon their utilization in the cycle of tricarboxylic acids (TCA). Although different molecules, such as lactate and various aminoacids, enter the TCA cycle at different levels, an average production of 28 moles ATP/moles of *A* nutrients has been assumed here. In this way, the utilization of A molecules can be expressed in terms of glucose equivalents allowing us 1) to normalize equation (16d) by $\mathbf{MW_G}$ and 2) to avoid writing specific mass-balance equations for each nutrient.**

**The gluconeogenesis pathway is described in our model by means of the rate $p_{11}$. This process costs a cell 2 moles of ATP/mole of *A* nutrients to build 1/2 mole of glucose, so that:**

$$ConsATP = 2 \cdot p_{11} \cdot \frac{MW_{ATP}}{MW_G} \qquad (18)$$

Thus, the overall rate of ATP production is modelled by the following equation:

$$ATP_TOT = ATP_nOx + ATP_Ox + ATP2 + ATP3 - ConsATP \qquad (19)$$

ATP is used by cells in a huge number of hierarchical biochemical reactions [24,25]. Globally, however, ATP production almost matches ATP utilization, at least when integrated over a critical time period [24,25]. Experimental data show nonetheless that during the life cycle of a cell its ATP content never falls to zero but rather tend to increase slightly [26]. Thus, some process of ATP storage and/or protection must exist to prevent complete utilization of ATP molecules. As it will be discussed later, this process of ATP preservation might allow a tumor cell to maintain a high energy status to proceed along the cell cycle. An example of storage mechanisms of high energy phosphoryl groups in the living tumor cells may include formation of phosphocreatine reservoirs [27]. Also, the high affinity binding of ATP molecules to the protein actin to form the polymerizable ATP-actin monomer might be a mechanism to preserve ATP from hydrolyzation [28,29].

Whichever is the mechanism exploited by cells to preserve ATP and increase their ATP pool during the cell cycle, we have addressed this point phenomenologically, i.e. by setting an appropriate rate of ATP consumption ($r_c$) whose value is given by the difference between the overall rate of ATP production (*ATP_TOT*) and an unknown rate $\beta$. This difference has been assumed to decrease for low overall ATP production rates:

$$\begin{cases} r_c = ATP_TOT - \beta & \text{if } ATP_TOT > ATP_St \\ r_c = ATP_TOT - \beta \cdot \dfrac{ATP_TOT}{ATP_St} & \text{elsewhere} \end{cases} \tag{20}$$

In this way, when the overall ATP production rate becomes critical the ATP is assumed to be mostly utilized by cells instead of being accumulated. The kinetics of the ATP pool have then been described by the following differential equation:

$$\frac{dATPp}{dt} = ATP_TOT - rc \tag{21}$$

The value of the rate $\beta$ can therefore be estimated by fitting the above equations to experimental data.

## 4. Modelling the cell cycle as a function of cell metabolism

The cell cycle comprises a series of programmed and highly coordinated events whereby a living cell duplicates its biochemical machinery and distributes it between daughter cells capable of carrying out the whole sequence again. A set of various proteins, collectively called cyclins, forms complexes with enzymes, the cyclin-dependent kinases. The complexes phosphorylate specific substrates at appropriate phases in the cell cycle, driving the cellular events necessary for progress from one phase to another. The result is a complex series of biochemical reactions with feedback regulatory loops which has been characterized and modeled in detail over the years [30].

The control of the progression along the cell cycle consists ultimately in the activation of kinases which phosphorylate substrates using ATP as the donor of high-energy phosphoryl groups [31,32]. It has been shown that tumor cells possess two main energetic checkpoints which are functions of the amount of available ATP [33,34]. These checkpoints regulate the progression of tumor cell during the G1 phase and their progression from the G2 to the M phase [33,34].

In Fig.2 and in the next paragraphs we show how we have modeled the energetic checkpoints and the cell cycle: this model bridges together cell metabolism and the cell cycle since the successful traversal of the energetic checkpoints depend on the amount of available ATP.

### *4.1 Energetic checkpoints along the cell cycle*

Newborn cells begin their life cycle by entering into a growth factor-dependent phase called G1. Insufficiency of growth factors and nutrients forces normal cells to leave the G1 phase and to become quiescent, a state called G0. Addition of nutrients induces G0 cells to re-enter the life cycle and to progress along the G1 phase and then irreversibly along the other three phases (S, G2 and M). The G1 phase is therefore subdivided into two further phases [35]: an initial subphase where cells are sensitive to variations in the environmental nutrient concentration (G1m) and a later subphase where cells are insensitive to deprivation of nutrients (G1p) [35]. Tumor cells are not believed to enter a true G0 phase upon nutrient depletion from the environment but nonetheless they stop progressing along the cell cycle [36]. Previous work has demonstrated that this behaviour of tumor cells is related to the intracellular concentration of their ATP pool which must be large enough to allow cells to overcome an energetic checkpoint placed within the G1 phase [33,34]. Once committed to go beyond this energetic checkpoint the successful passage of cells through G2 into mitosis (M phase) is also conditional upon maintenance of a critical ATP content sufficient to satisfy a second energy-sensitive checkpoint [33,34]. The G1 energetic checkpoint appears to be more sensitive to ATP content than the second checkpoint.

Reduction of the ATP pool by pharmacologic treatment reveals that tumor cells can function with as little as 70% of normal ATP content without significant alteration of their growth [34]. Reduction of ATP concentration below 15% of normal levels results in cell death. In between these two values cells stop growing [34]. Thus, ATP levels in the range 15-70% of the normal concentration determine the thresholds of the energetic checkpoints.

We have modelled these observations by setting two ATP thresholds (Fig2). It should be noted that the difference between the rates of ATP production (*ATP_TOT*) and ATP consumption ($r_c$) is always positive (see the rate $\beta$ in equation 18) or null. Thus, the ATP pool is always increasing or constant throughout the cell cycle. We have also assumed a cost for cells whose ATP pool is high enough to overcome the thresholds, which is an amount of ATP equivalent to the threshold levels. This amount is instantaneously consumed thus resetting the ATP pool of a cell to the minimum level needed for survival.

Thus the duration of the cell cycle depends on the time needed to produce enough ATP to overcome the energetic checkpoints and therefore on cell metabolism. The S and the M phases have been assumed to last a fixed time span (Fig.2). A further constraint has been set on the minimum duration of the G2 phase (Fig.2).

*4.2 Increase in cell volume and in the number of mitochondria*

Most ATP is produced by tumor cells through the oxidative phosphorylation pathway whose key final steps occur into mitochondria. Mitochondria proliferate by fission into cells during their life cycle. Thus, the amount of produced ATP should be proportional to the number of mitochondria present in a cell. This number may vary considerably in different cell types ranging from 83 to 677 mitochondria per cell [37].

Since cells increase volume and surface area during their life cycle the amount of nutrients taken up by cells would also increase. Moreover, it has been shown that the cell volume is quasi-linearly related to the number of internal mitochondria during the cell cycle [38].

The uptake of glucose from the surrounding environment is described by rates $v_{1p}$ and $v_{1m}$ and its consumption is ultimately modelled by rates $g_1$ and $g_2$. Thus, the net balance of stored glucose into cells, independently of glucose form (i.e. as glucose or as G6P or as glycogen), is a linear combination of the above rates and its time-dependent variations follow the differential equation:

$$\frac{dNetG}{dt} = (v_{1p} - v_{1m}) - (g_1 + g_2) \qquad (22)$$

We have then assumed that a fraction of stored glucose is used by cells as a source of carbon to build up their structures and hence to increase their size (mass and volume). Building internal structures has a cost in terms of energy and this cost is represented by the rate $r_c$ of ATP utilization, that, as described above, is intended to globally models the energy needed to carry out protein and DNA synthesis and to maintain cell functions (e.g. ion transport). Obviously, a cell cannot exceed a maximum volume size. Thus, the equations describing the time-dependent variations in cell volume, and its boundary limit, are parametrized as follows:

$$\begin{cases} \dfrac{dVolume}{dt} = \alpha \cdot NetG \cdot r_c \\ \lim\limits_{t \to \infty} Volume(t) = Vol_{\max} \end{cases} \qquad (23)$$

where $\alpha$ is a parameter which assumes positive values. At all times, the number of mitochondria ($M$) increases proportionally to cell volume up to a maximum value ($M_{\max}$):

$$M(t) = Volume(t) \cdot \frac{M_{\max}}{V_{\max}} \qquad (24)$$

and therefore the number of mitochondria is approximated by a real number. Rates $r_2$, $g_2$, $r_3$, $p_{22}$ and $ATP_St$ that describe processes occurring within mitochondria have then been normalized with respect to the fraction of mitochondria $M(t)/M_{\max}$ present within a cell at all times.

*4.3 Cell division at mitosis and partitioning of mitochondria between daughter cells*

At the end of the M phase a cell divides and distributes its biochemical machinery, chromosomes and organelles between the two daughter cells. The partitioning of subcellular organelles has been studied in eukaryotic cells and found to follow roughly a binomial distribution [39,40]. However, deviation from a pure binomial distribution has been observed [40]. Tumor cells within a cell population are known to show variable volume sizes, a variable DNA content (due to imperfect formation of the mitotic spindle) and a variable content in mitochondria [41]. In the latter case, it is also known that mitochondria can be retained by the mother cell upon actin-mediated intracellular migration to protected compartments [39]. Complex yet unknown mechanisms underlay the partitioning of mitochondria during cell division. As a consequence it is difficult to find a theoretical basis to justify the choice of one particular statistical distribution. For this reason we have taken a rough approximation by randomly choosing the volume size of daughter cells from a uniform distribution between the two limit volume sizes $Vol_{\min}$ and $Vol_{\max}$, where $Vol_{\max}$ has been defined above. To set the minimum volume (see Tab.1) we noted that $Vol_{\min}$ should not assume values below the average volume size of the cell nucleus. As soon as the volume of daughter cells is determined, the number of mitochondria is computed using equation (24).

## 5. Simulations

*5.1 Cell metabolism: outputs of the simulations vs. experimental data*

**The present simulator takes into account the parameters listed in Tab.1. For some of these parameters experimental values are available in the literature. However, as described in the previous sections, some metabolic processes have been approached phenomenologically and the average values for the phenomenological parameters have been estimated by fine tuning the whole model at the single cell level to experimental metabolic data (Tab.1). To this end one cell was repeatedly allowed to grow for just two cell-cycles. The program was then used to simulate the behaviour of a large cell population for an increased time span and average matabolic rates were computed and compared to actual experimental data.**

A population of 500 cells has been simulated using the parameters given in Tab.1. Since the initial conditions are the same for all cells, the simulator has been allowed to run for a total time span of $5x10^6$ s to obtain a completely desynchronized cell population (see also Fig.6). Experimental data on cell metabolism are in general measured on desynchronized cell populations cultured *in vitro*.

Metabolic rates were averaged over further 80000 s of simulation time. These are the rate of overall glucose uptake by cells from the environment (given by the difference $v_{1p}$-$v_{1m}$, see equations 3b and 3c), the rate of lactic acid production (equation 17), the overall rate of ATP production through the oxidative phosphorylation (equations 16b-d), the rate of ATP production through the anaerobic glicolysis (equation 16a) and the rate of oxygen consumption (equation 8). It should be noted that these rates are computed at different levels within the hyerarchy of the algorithm, the rate of glucose uptake by cells being at the very first level and the rate of oxygen consumption being at the core of the algorithm (see e.g. Fig.1). **The numerical simulation of the metabolic rates at different levels along the metabolic network was carried out to check the correct implementation of the network.**

Computed values for the above rates are given in Tab.2. These values are in good agreement with experimental data, **also demonstrating the robustness of the model in long-lasting simulation of large cell populations**.

*5.2 On the interplay between glucose uptake and oxygen consumption: the Pasteur and the Crabtree effects*

**The major goal of the present work is to develop a model capable of reproducing *in silico* the main aspects of cell metabolism and growth at the cell population level. The successful simulation of these basic behaviours represents in turn a check of the validity of the model.**

One of the best-recognized "universal" patterns of the metabolism of living organisms is the so-called Pasteur effect whereby cells consume glucose at higher rates upon reduction of environmental oxygen [17]. The model can simulate the Pasteur effect by means of the **phenomenological** parameter $h$ defined in equation (2e) which tunes the maximum rate of glucose uptake by cells as a function of the intracellular oxygen concentration (Fig.3).

Conversely, when tumor cells are exposed to low environmental glucose concentrations they consume more oxygen, or equivalently when they are grown in a high glucose environment they consume less oxygen [21,22]. This inhibition of respiration by glucose, known as the Crabtree effect, has been observed in several tumor highly glycolytic cells and tissues [42]. The multifactorial Crabtree effect, which to our knowledge has not yet a definitive molecular explanation, can also be appreciated in our model for glucose concentrations below standard physiologic ones (Fig.4). This is due to the ATP sensor described in Section 2 which tunes the rates $r_3$, $p_{22}$ and $p_{11}$ of *STORE* glycogen and non-glucose molecules utilization, respectively (see also Fig.1). At low concentrations of glucose, the ATP production rate falls below the standard value (*ATP_St*) and turns on the ATP sensor. Additional *STORE* molecules as well as non-glucose molecules in the *A* compartment are then catabolized resulting in a transient overshooting of ATP production and oxygen consumption.

*5.3 Cell metabolism determines the length of the cell cycle*

Fig.5 shows the quantitative relationships between the amount of the intracellular ATP pool and the duration of the individual phases of the cell cycle. Using the parameters given in Tab.1 the overall length of the cell cycle results on average of 22.2 hours. Because of random division of the cell volume at mitosis and therefore of the number of mitochondria, where oxydative phosphorylation takes place, the kinetics of the ATP storage may vary among cells. As a consequence, the duration of the cell cycle may also vary among cells within the same population. This leads to the desynchronization of the cycle of the whole population, an effect that has been directly measured for tumor cells grown in vitro and which allowed us to test the global behaviour of the simulator.

*5.4 On the desynchronization of the cell cycle*

A population of 1000 cells has been simulated using the parameters given Tab.1, and the number of cells in each phase of the cell cycle has been sampled every 5000 s. Since the initial conditions are the same all cells are initially synchronized. Upon random division, synchronization is progressively lost and complete desynchronization is observed after approximately $1.2x10^6$ s (Fig.6).

Tab.3 compares the percentage of virtual cells in the different phases with that measured in our laboratory for various cell lines using standard cytofluorimetric techniques [6].

Recently, the desynchronization rate of tumor cells has been measured by cytofluorimetry using an elegant biparametric technique [43]. This technique allows the experimenter to label all the cells which are in the S phase at some time during the cell cycle. Since the cells are labelled they can be identified and followed in time during the cell cycle. A plot of the amount of cells wich re-enter together the S phase as a function of time gives a direct measurment of the speed of convergence towards the asynchronous distribution of the cell population [43]. The same strategy can be followed numerically using our simulator. Fig.7 (upper panel) shows the results of two simulations carried out with different choices of the parameter $\beta$ which tunes the kinetics of ATP storage and hence of the cell cycle. To compare the simulated outputs with experimental data the data given in [43] have also been drawn in Fig.7 (bottom panel).

## 6. Conclusion and Outlook

We wish to stress again that the present model is not intended to be a rigorous, exact model of some specific cellular line, but is only meant **to realistically simulate the metabolism, growth and proliferation dynamics of large tumor cell populations.** The basic usefulness of this approach is justified by the down-to-earth observation that while cells may differ in a large number of specific biochemical paths, they still follow some common, universal rules (this is very foundation of all kinds of biological classification, after all). In other words we rely on that kind of averaging that has been so successful in statistical mechanics, where the behavior of large complex systems like real magnets is realistically reproduced by simple systems composed by elementary parts like the Ising and Heisenberg magnets [44], and indeed here we want to model the behavior of large populations of cells (> $10^6$ cells) for long time spans (up to months) [3]. As a consequence, in our mesoscopic approach several details of cell metabolism and of the cell cycle have been parametrized and averaged. In a sense, we try to apply to cell biology the methods that have been so successful in statistical mechanics, and set up a kind of "statistical cell biology".

We have incorporated in the model two basic aspects of cell biology: the kinetics of glucose uptake and of oxygen consumption by tumor cells and the interplay between these biochemical parameters and the cell cycle. Experimental observations point to glucose and oxygen as the key molecules to explain why the cells colonizing the inner parts of a solid tumor become quiescent and eventually die by necrosis [14,45-47], therefore the model checks have been concentrated on the interplay of oxygen and glucose metabolism.

**One important assumption in the present model is that tumor cells consume a fixed fraction of nutrients to sustain growth and proliferation (see chapter 3.2), an assumption that allowed a considerable simplification of the metabolic model which otherwise should have included all the metabolic steps from the formation of G6P to lactate and pyruvate and the further steps of the cycle of tricarboxylic acids. The fact that tumor cells must consume a fraction of nutrients to sustain proliferation is not unreasonable, but whether this fraction is fixed or not may be questionable. We think that this hypothesis might be verified experimentally by keeping cells under different nutrient conditions and by measuring the corresponding rates of nutrient utilization. If the fraction of nutrients used for growth turned out to be non-constant, the simulator might be further refined by introducing the observed utilization rates.**

The simulations show that the model approximates quantitatively some of the available experimental data, at least as far as glucose metabolism, oxygen consumption, ATP and lactic acid production by tumor cells and the kinetics of the cell cycle are concerned. It is worth noting that in these simulations our model uses as far as possible parameters obtained from experimental observations or **tuned to fit experimental data**. **By tuning parameters to experimental data, the behaviour of specific cell types can be reproduced using the very same simulation framework. This framework allows a flow of information to occur from single molecules to the whole cells through cell metabolism and the two-way connection between metabolism and the cell cycle. Information propagates then to the cell population as the behaviour of individual cells in response to the environmental changes influences the collective behaviour** **of the entire population. For example, the desynchronization of the cell cycle of the simulated population shown in Fig.6 and Fig.7 – resulting from the stochastic processes considered at the cellular and subcellular level (e.g. random partitioning of the mitochondria) – realizes the transfer of information between different spatiotemporal scales, and reproduces quantitatively the experimental data.**

We noted above that numerical simulations of large cell clusters cannot include too many details if we want to keep the simulation time and memory requirements from blowing up, and that a model should maintain a high plasticity (limited network structure and ability to adjust parameters) to simulate the behavior of different cell lines. On the other hand we would like to include as many details as possible in order to provide an accurate simulation of cells and their enviroment. How deeply should we push our description of tumor cell metabolism and of the cell cycle? We do not know yet, the answer can only come from better experimental data and from further numerical tests, but at the moment we feel that the minimal model that we have developed compares very favorably with many different existing data in three related parts of cell biology: metabolism, growth and proliferation.

## Acknowledgments

This work has been supported by grants from the Istituto Nazionale di Fisica Nucleare (Italian National Institute for Nuclear Physics), VIRTUS project, Section V. The authors wish to thank Dr. Chiara Dalla Pellegrina and Dr. Giorgia Padovani for their technical help in performing cytofluorimetric assays. Prof. Giancarlo Andrighetto is gratefully acknowledged for helpful comments and suggestions on various parts of this work. We also wish to thank the anonymous referees for their thoughtful reading of the manuscript and for their constructive criticism.

**Appendix: diffusion and absorption times for oxygen.**

Take any chemical species that diffuses from the outside to the inside of a cell and assume that its diffusion constant in the membrane is much smaller than the diffusion constant in the cytoplasm, so that its concentration is nearly uniform inside the cell. Now let $\phi_{in}$ and $\phi_{out}$ be the concentrations inside and outside the cell, then the diffusion current from the outside to the inside is

$$J = D_m \frac{\phi_{out} - \phi_{in}}{h}$$

where $h$ is the membrane thickness and $D_m$ is the diffusion coefficient in the membrane. If both diffusion and absorption are present, the concentration change in a time interval $\Delta t$ is

$$V\Delta\phi_{in} = D_m \frac{\phi_{out} - \phi_{in}}{h} S\Delta t - \lambda V \phi_{in} \Delta t$$

where $\lambda$ is the absorption rate per unit volume, $S$ is the cell surface, and $V$ is the cell volume. From the last equation we get

$$\frac{d\phi_{in}}{dt} = D_m \frac{\phi_{out} - \phi_{in}}{h} \frac{S}{V} - \lambda\phi_{in}$$

so that the time constant is

$$\tau^{-1} = D_m \frac{S}{hV} + \lambda$$

For an approximately spherical cell we have, $\frac{S}{V} \approx \frac{3}{r}$ where $r$ is the cell radius, and then

$$\tau^{-1} \approx D_m \frac{3}{hr} + \lambda$$

In the case of oxygen we take the values (see also Tab.1)

$\lambda \approx 10^{-7}\ s^{-1}$

$D_m \approx 1.5 \cdot 10^{-11}\ m^2 s^{-1}$ (see ref.[23])

$r \approx 5 \cdot 10^{-6}\ m$

$h \approx 10^{-7}\ m$

and then we find that the time constant for simple absorption is

$\tau_A \approx 10^7 s$

while when we turn on diffusion this changes to

$\tau_{AD} \approx 10^{-2} s$

In conclusion, when diffusion is present we can neglect absorption, and we can assume that the outside and the inside concentrations are actually the same.

**Figure captions**

**Fig.1** – Scheme of the biochemical network that models cell metabolism. Variables within circles represents molecular species and are therefore expressed in units of concentration or mass. Symbols are as follows: $G_{ex}$=external glucose concentration, $G_{in}$=intracellular glucose concentration, *G6P*=glucose-6-phosphate concentration, *STORE*=mass of glucose stored in the form of glycogen, *Ac.Lat*=lactic acid concentration, *A*=mass of nutrients other than-glucose, $O_2$=oxygen concentration, *ATPp*=pool of ATP molecules concentration. The rates of conversion of considered molecular species have been represented by squares. The dotted circuit represents the action of the oxygen sensor ($SensO_2$) on the indicated rates. The sensor is roughly the difference between available oxygen molecules and those consumed during ATP production (see equation 10). The dashed circuit shows the action of the ATP sensor (*SensATP*) on the indicated rates. The ATP sensor monitors at each time the amount of ATP produced through the oxidative phosphorylation of glucose molecules and compare this rate (*ATP_OX*) with a standard ATP production rate (see equations 7 and text for details).

**Fig.2** – Modelling cell cycle kinetics as a function of cell metabolism. The pool of ATP molecules (*ATPp*) increases in time as a function of cell metabolism which takes place within an increasing number of mitochondria. Mitochondria divide by fission with kinetics proportional to cell volume. The cell volume is hypothesized to increase in time because of the coupling between storage unused glucose and energy consumption. Nutrient uptake is also influenced by the growth of the cell surface. The cell cycle is subdivided into the four canonical phases G1, S, G2 and M. In addition, the G1 phase is further subdivided into two phases (G1m and G1p) on the basis of the existence of an ATP threshold (ATP threshold 1). One additional ATP threshold (ATP threshold 2) is placed at the border between the G2 and M phases. The existence of ATP thresholds stems from current experimental evidence. Cells within the G1m phase are sensitive to nutrients and can proceed toward the G1p phase if their ATP content overcomes the ATP threshold 1. However, the cost that cells must pay to enter the G1p phase is an amount of ATP proportional to the ATP threshold that is "instantaneously" consumed by cells. Cells then leave the G1p phase when their ATP content ideally matches that of the ATP threshold 1. As a consequence the duration of both G1m and G1p phases is variable and depends on cell metabolism. The S and M phases have been assumed to last a fixed time span which is indicated in the figure. The G2 phase, instead, is characterized by a minimum time span which is required by cell to complete molecular process to prepare them to enter mitosis. These processes require the availability of a minimum amount of ATP defined by the ATP threshold 2.

**Fig.3** – Simulating the Pasteur effect. A desynchronized population of 500 cells (obtained by running the algorithm for a total of $5x10^6$ s virtual time) was allow to "grow" for 80000 s at time steps of 1 s. The rate of glucose uptake by each cell was sampled every 1000 s. The rates were then averaged (continuous line) and the standard deviation (dotted lines) calculated. At the time indicated by the arrow the environmental oxygen concentration was instantaneously decreased to levels corresponding to 10% of the standard level (standard concentrations are given in Tab.1). Cells react by increasing their rate of glucose uptake (the Pasteur effect).

**Fig.4** – Inhibition of respiration by glucose: an analog of the Crabtree effect. Cells were "prepared" as explained in the caption of Fig.3. At the time indicated by the arrow, the environmental glucose concentration was decreased to levels corresponding to 10% of the standard level (standard concentrations are given in Tab.1). In this case, the mean rate of oxygen consumption (continuous line) and standard deviation (dotted lines) have been computed. By lowering the environmental glucose concentration the rate of oxygen consumption increases. Glucose, therefore, inhibits respiration (Crabtree effect).

**Fig.5** – Relationship between the time-dependent variations of the ATPp concentrations and the phases of the cell cycle for a single cell. *Upper panel*: the phases of the cell cycle are represented by numbers as follows: 1=G1m, 2=G1p, 3=S, 4=G2 and 5=M. *Bottom panel*: the pool of ATP molecules is sharply reduced upon overcoming of the ATP thresholds (see also Fig.2). **In this simulation, as well as in the following simulations, the thresholds have been set at $4.128 \times 10^{-15}$ and $2.408 \times 10^{-15}$ Kg ATP.** The figure shows that the overall duration of the cell cycle is approximately 22.2 hours. However, the length of the cell cycle is expected to vary as a consequence of unequal division of the cell at mitosis (see text for details).

**Fig.6** – Cell cycle analysis of a population of 1000 cells. Cells starts living at the same instant at the beginning of the G1m phase. The algorithm was run for $2 \times 10^6$ s using timesteps of 1 s. Every 5000 s the number of cells within each phase of the cell cycle was computed. The initial imposed synchronization of the cell cycle is lost after approximately $1.2 \times 10^6$ s of computer time.

**Fig.7** – Convergence of cells towards the asynchronous state. *Upper panel*: a population of 1000 cells was grown *in silico* for $5 \times 10^6$ s at time steps of 1 s and allowed to desynchronize (see also Fig.6). After that time, cells in the S phase were labelled and the population was simulated for the indicated time span. Every 3600 s the percentage of labelled cells occupying the S phase was computed. Simulations were carried out for two values of the parameter $\beta$ which tunes the amount of ATPp molecules stored by cell per unit time and hence the length of the cell cycle (continuous line: $\beta = 8.2 \times 10^{-20}$ kg $s^{-1}$; dashed line: $\beta = 5.2 \times 10^{-20}$ kg $s^{-1}$). *Bottom panel*: experimental data on the speed of convergence towards the asynchronous state for two human tumor cell lines, IGROV1 and MOLT4. Data have been taken from ref.[43].

In comparing the two figures one should note the different time scale on the horizontal axis.

**TABLE 1 Model parameters**

| Symbol | Value | Units | Meaning | References |
|---|---|---|---|---|
| $G_{ex}$ | 0.9 | kg m$^{-3}$ | environmental standard glucose concentration | [17] |
| $O_2$ | 3.392 10$^{-2}$ | kg m$^{-3}$ | environmental standard oxygen concentration | [17] |
| *ATP_St* | 2.017 10$^{-18}$ | kg s$^{-1}$ | rate of ATP production under standard conditions | **<u>derived</u> from fitting the model to data in [8]** |
| *A* | 3.6 | kg m$^{-3}$ | maximum amount of nutrients other than glucose | **derived from fitting the model to data in [8]** |
| *STORE* | 0.18 10$^{-16}$ | kg | maximum amount of *STORE* molecules | [8] |
| $Vol_{max}$ | 2.48 10$^{-16}$ | m$^3$ | maximum cell volume | [a][41] |
| $Vol_{min}$ | 0.79 10$^{-16}$ | m$^3$ | minimum cell volume | **[b]derived from data in [49]** |
| $M_{max}$ | 100 | | maximum number of mitochondria per cell | [37] |
| $V_m$ | 2.0 10$^{-9}$ | kg s$^{-1}$ m$^{-2}$ | maximum rate of glucose transport | [48] |
| $K_{m1}$ | 0.2704 | kg m$^{-3}$ | Michaelis-Menten constant of glucose transport | [48] |
| $V_{m2}$ | 1.2 10$^{-19}$ | kg s$^{-1}$ | maximum rate of glucokinase activity | [19] |
| $V_{m22}$ | 1.2 10$^{-18}$ | kg s$^{-1}$ | maximum rate of hexokinase activity | [19] |
| $K_{m2}$ | 1.8 | kg m$^{-3}$ | Michaelis-menten constant for glucokinase | [19] |
| $K_{m22}$ | 1.8 10$^{-2}$ | kg m$^{-3}$ | Michaelis-Menten constant for hexokinase | [19] |
| $coeffg_1$ | 7.5 10$^{-3}$ | s$^{-1}$ | fraction of G6P molecules entering the anaerobic glycolysis per unit time | **derived from fitting the model to data in [8]** |
| $coeffg_2$ | $1.08\ 10^{-3}$ | $s^{-1}$ | fraction of G6P molecules entering the oxidative phosphorylation per unit time | **derived from fitting the model to data in [8]** |
| $coeffg_3$ | $6.7\ 10^{-4}$ | $s^{-1}$ | fraction of G6P molecules converted into STORE molecules per unit time | **derived from fitting the model to data in [8]** |
| $coeffr_1$ | $3\ 10^{-20}$ | $s^{-1}$ | fraction of STORE molecules entering the anaerobic glycolysis | **derived from fitting the model to data in [8]** |
| $\boldsymbol{coeffp_{11}}$ | **1** | $\mathbf{s^{-1}}$ | **fraction of A molecules entering the STORE compartment** | **derived from fitting the model to data in [8]** |
| $K_a$ | $5.4\ 10^{-2}$ | $kg\ m^{-3}$ | constant for the homeostatic loop controlling glucokinase and hexokinase activity | **derived from fitting the model to data in [8]** |
| $K_c$ | $3.6\ 10^{-17}$ | kg | constant for the homeostatic loop controlling STORE molecules consumption | **derived from fitting the model to data in [8]** |
| $K_d$ | $1.8\ 10^{-2}$ | $kg\ m^{-3}$ | constant for the homeostatic loop controlling *A* molecules consumption | **derived from fitting the model to data in [8]** |
| $\alpha$ | $2.18\ 10^{-16}$ | $m^3\ s\ kg^{-2}$ | phenomenological constant for cell volume increase | **derived from fitting the model to data in [38]** |
| $\beta$ | $8.2\ 10^{-20}$ | $kg\ s^{-1}$ | phenomenological constant for ATP storage | **derived from fitting the model to data in [26]** |

[a]The cell volume may vary considerably in size for different cells. Here, the volume size of human neutrophils has been considered.

[b]Considering that the cell nucleus occupies between 4% and 14% of cell volume [49].

**TABLE 2 Comparison between simulated metabolic parameters and experimental ones**

| Rates | Simulations (mean±SD) | [a]Experiments (mean±SD) |
|---|---|---|
| Glucose uptake | [b]$1.89 \times 10^{-19} \pm 3.1 \times 10^{-20}$ | $2.54 \times 10^{-19} \pm 1.8 \times 10^{-20}$ |
| Lactate production | $3.78 \times 10^{-19} \pm 3.1 \times 10^{-20}$ | $3.91 \times 10^{-19} \pm 8.1 \times 10^{-20}$ |
| ATP through oxidative phosphorylation | $1.98 \times 10^{-18} \pm 8.3 \times 10^{-19}$ | $3.78 \times 10^{-18}$ |
| ATP through glicolysis | $1.06 \times 10^{-18} \pm 1.3 \times 10^{-19}$ | $1.14 \times 10^{-18} \pm 2.3 \times 10^{-19}$ |
| Oxygen consumption | $2.50 \times 10^{-20} \pm 1.1 \times 10^{-20}$ | $4.78 \times 10^{-20} \pm 1.0 \times 10^{-20}$ |

[a]Experimental data were taken from ref.[8].

[b]All values are expressed in kg $s^{-1}$.

**TABLE 3 Percent cells in the various phases of the cell cycle: simulation results vs. cytofluorimetric experiments**

| Cell type | G0/G1 | S | G2/M |
|---|---|---|---|
| Simulated | 54.9 ± 1.5 | 28.4 ± 1.4 | 16.6 ± 1.2 |
| MOLT3 (human T lymphoblastoid cell line) | 54.4 ± 2.2 | 27.5 ± 5.8 | 16.4 ± 1.7 |
| Jurkat (human CD5+ T lymphoblastoid cell line) | 59.0 ± 1.5 | 21.5 ± 0.2 | 15.8 ± 0.2 |
| [a]Proliferating normal human T lymphocytes | 62.6 ± 2.3 | 18.7 ± 0.9 | 18.7 ± 0.4 |
| CACO2 (human colon carcinoma cell line) | 40.1 ± 3.4 | 23.0 ± 0.4 | 33.0 ± 2.8 |
| WHEI 164 (mouse fibrosarcoma cell line) | 67.0 ± 2.9 | 11.0 ± 0.9 | 18.7 ± 4.6 |

[a]Lymphocytes drawn from healthy donors stimulated in vitro with phytohemagglutinin from *Phaseolus vulgaris* [6].

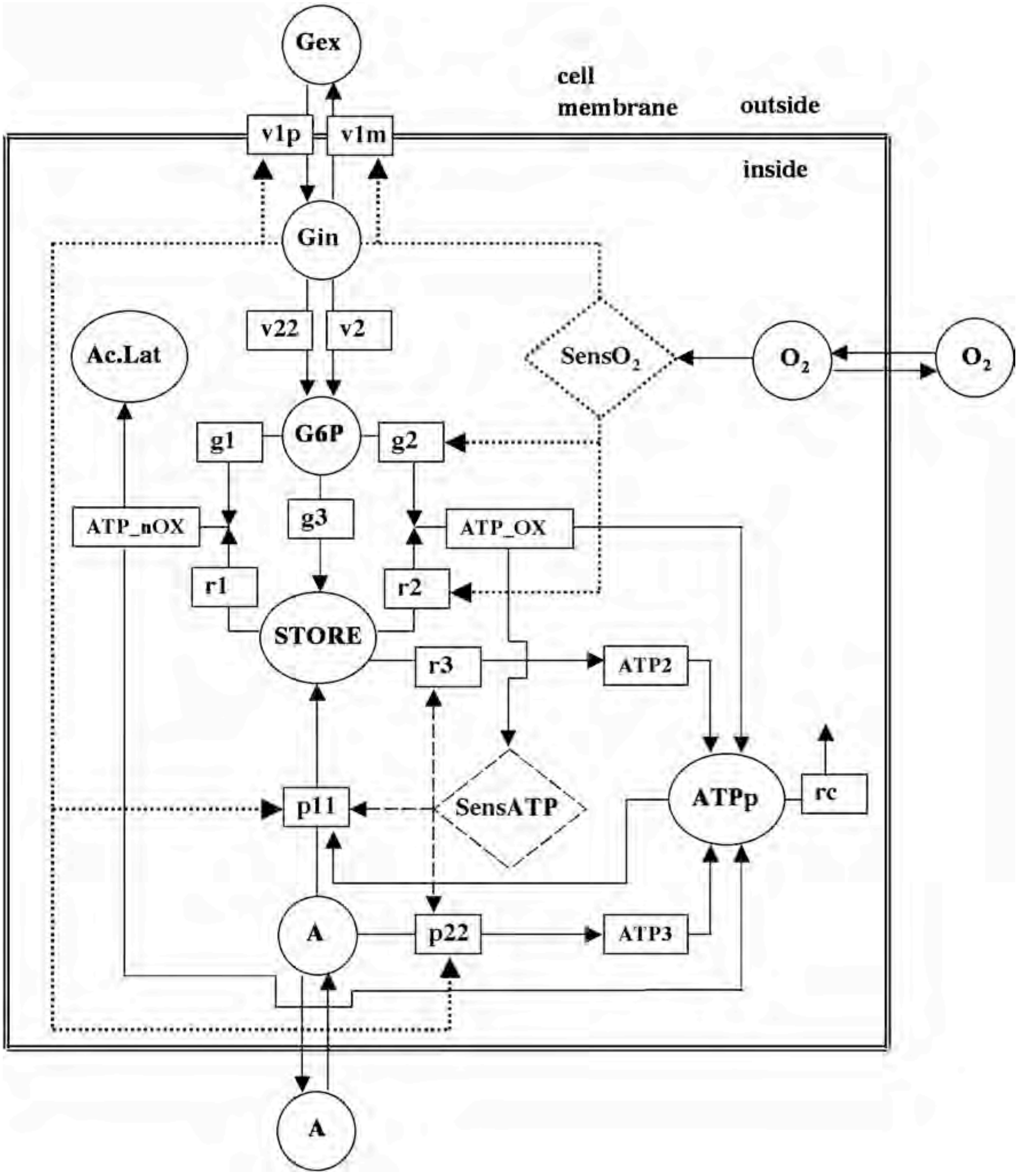

Fig.1 R.Chignola and E.Milotti

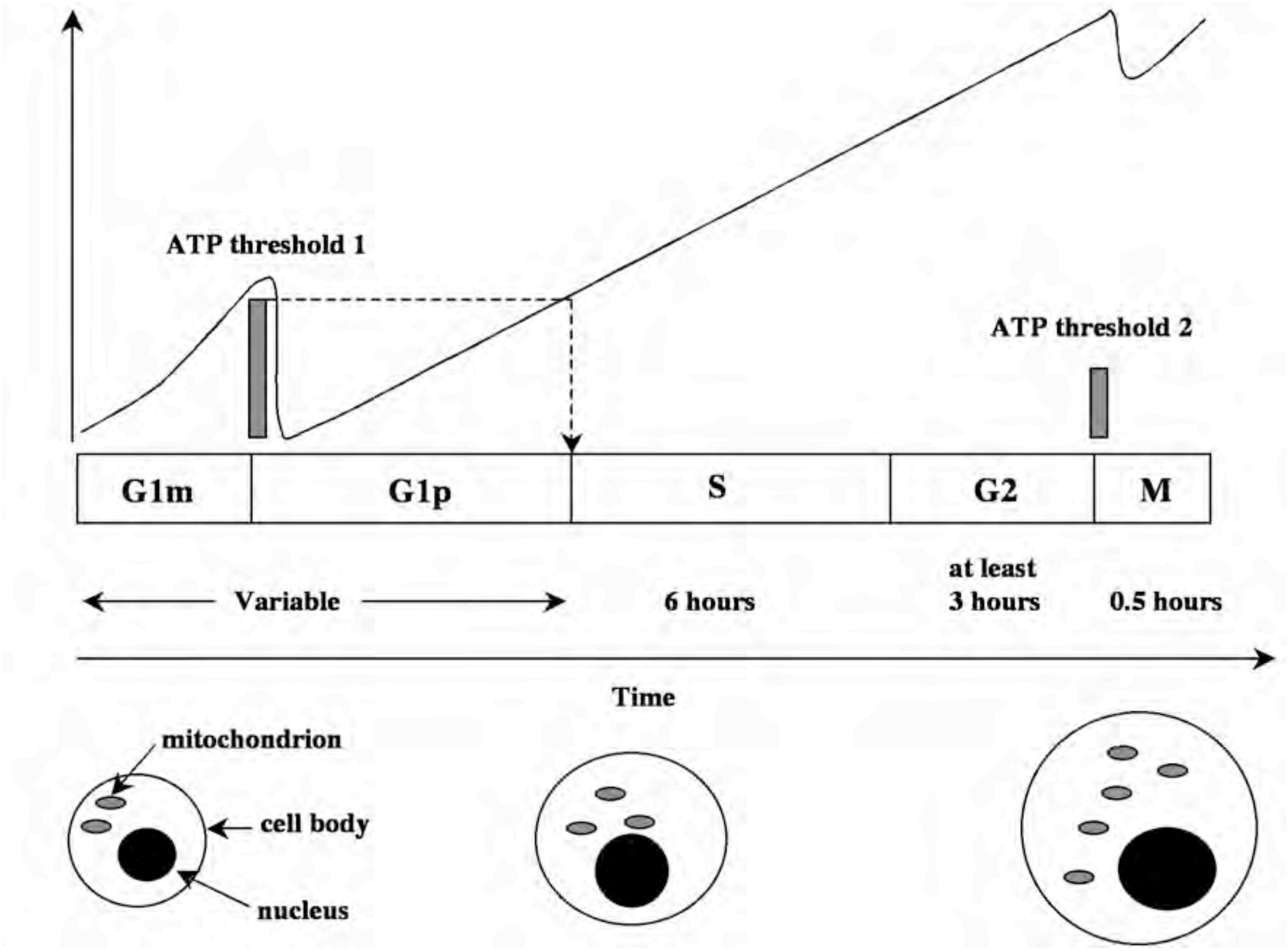

Fig.2 R.Chignola and E.Milotti

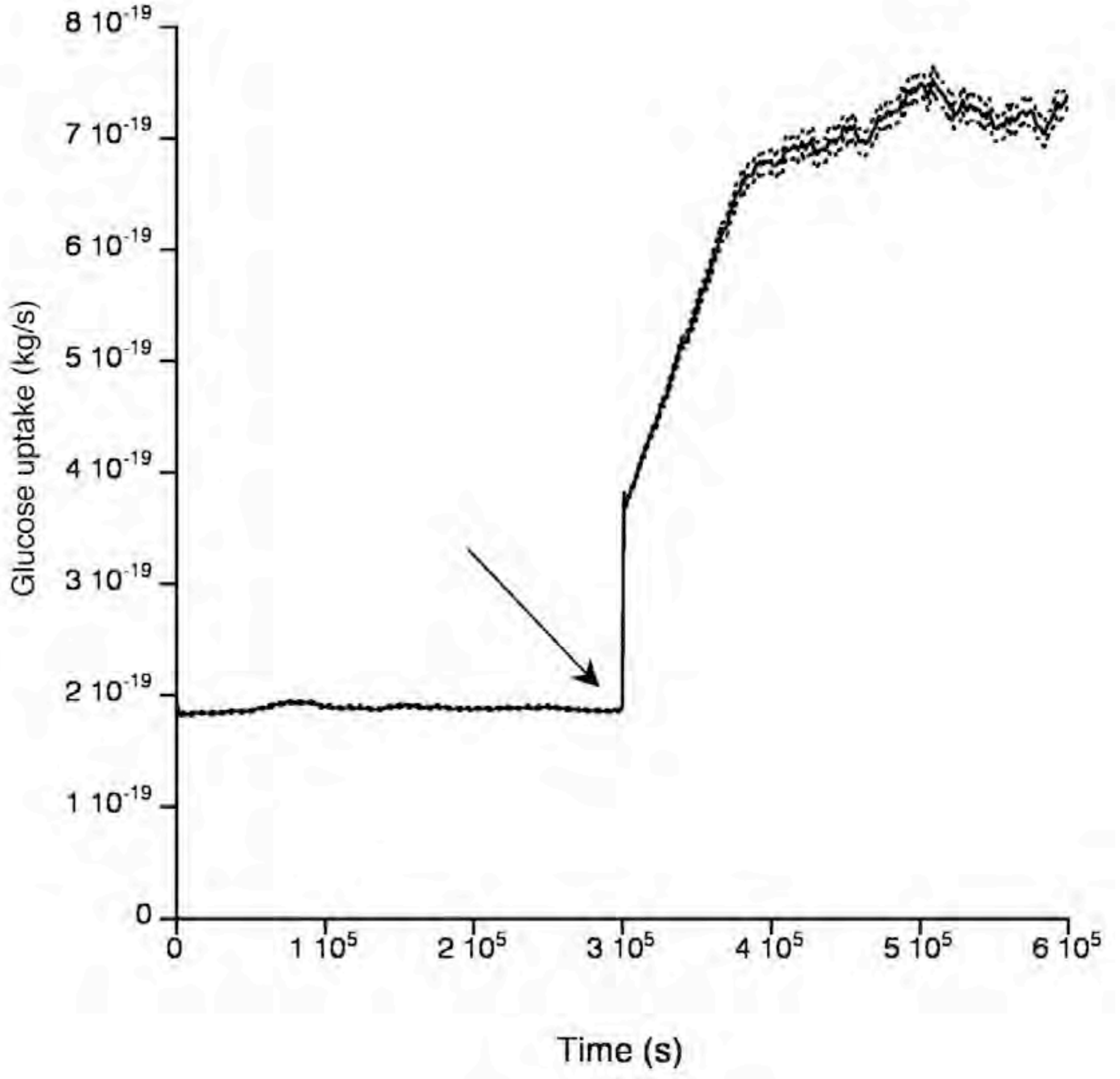

Fig.3 R.Chignola and E.Milotti

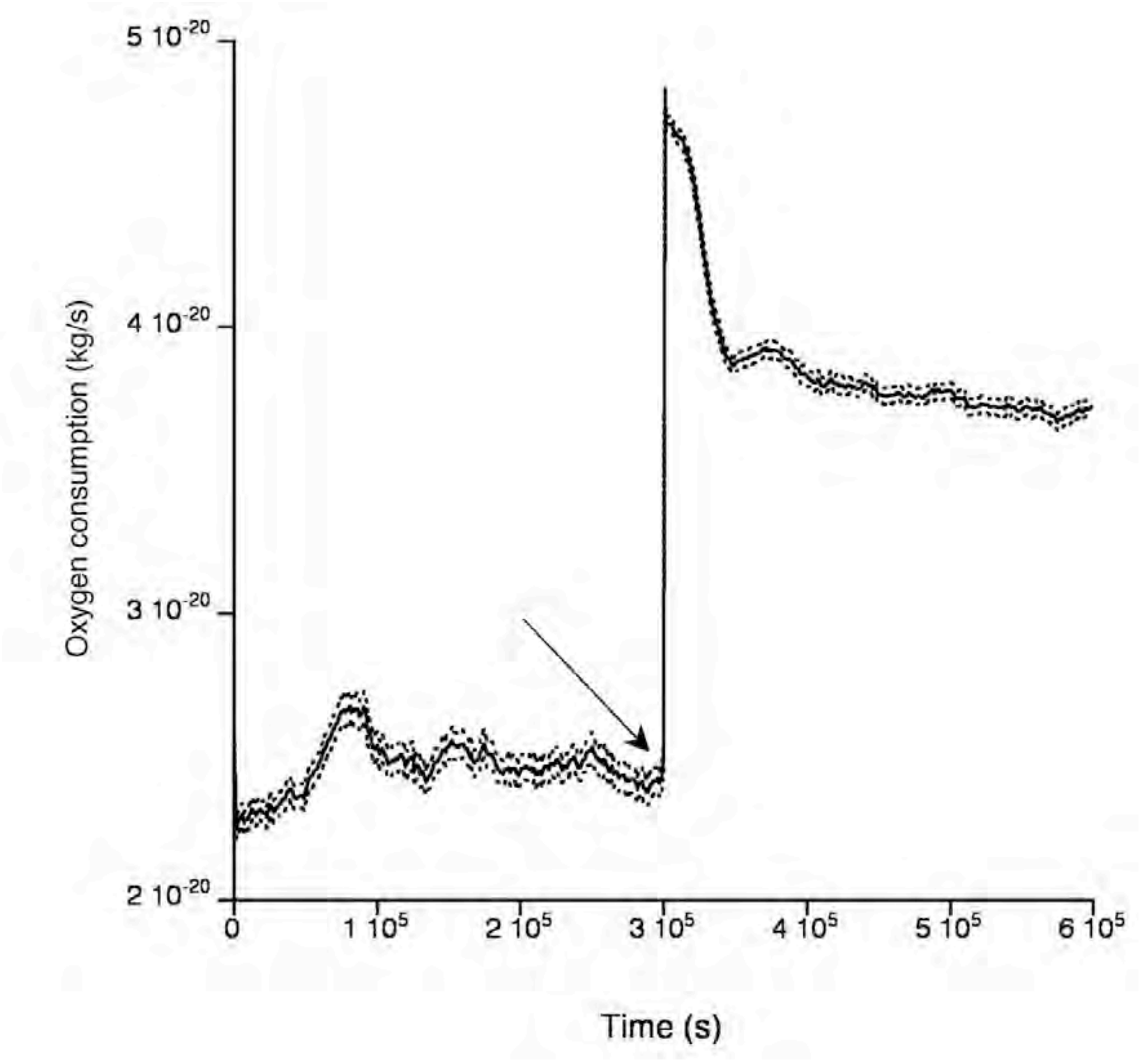

Fig.4 R.Chignola and E.Milotti

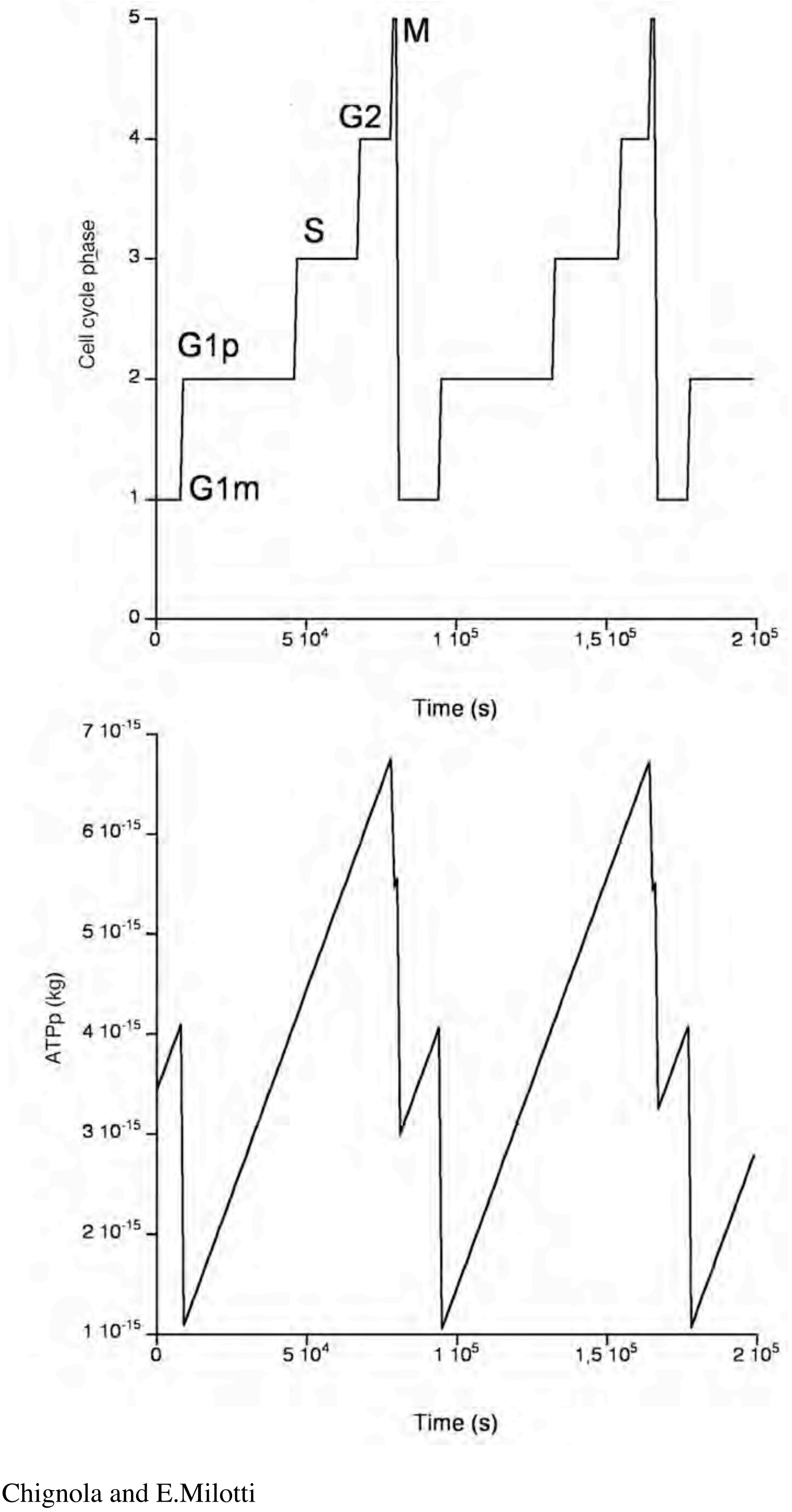

Fig.5 R.Chignola and E.Milotti

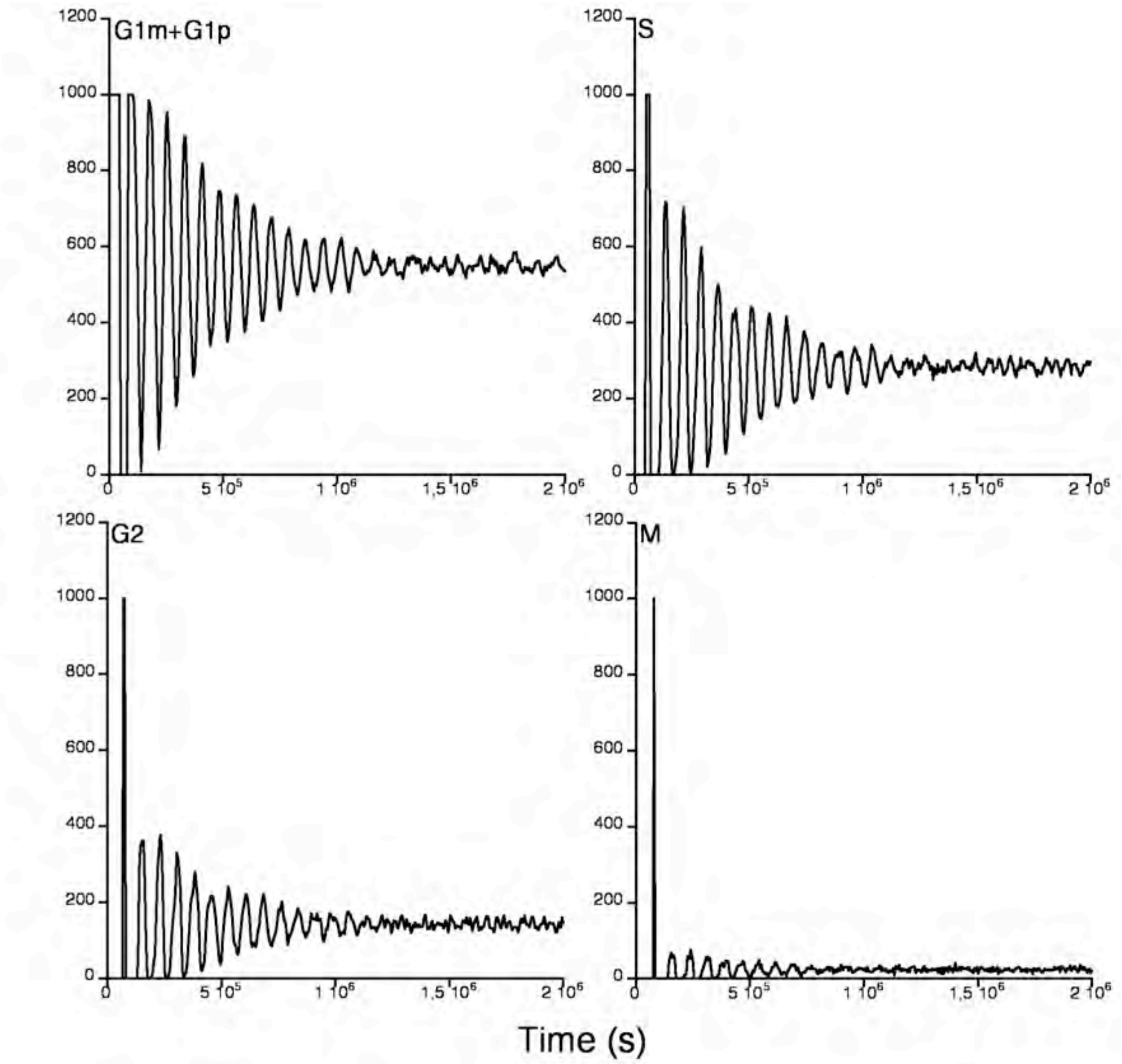

Fig.6 R.Chignola and E.Milotti

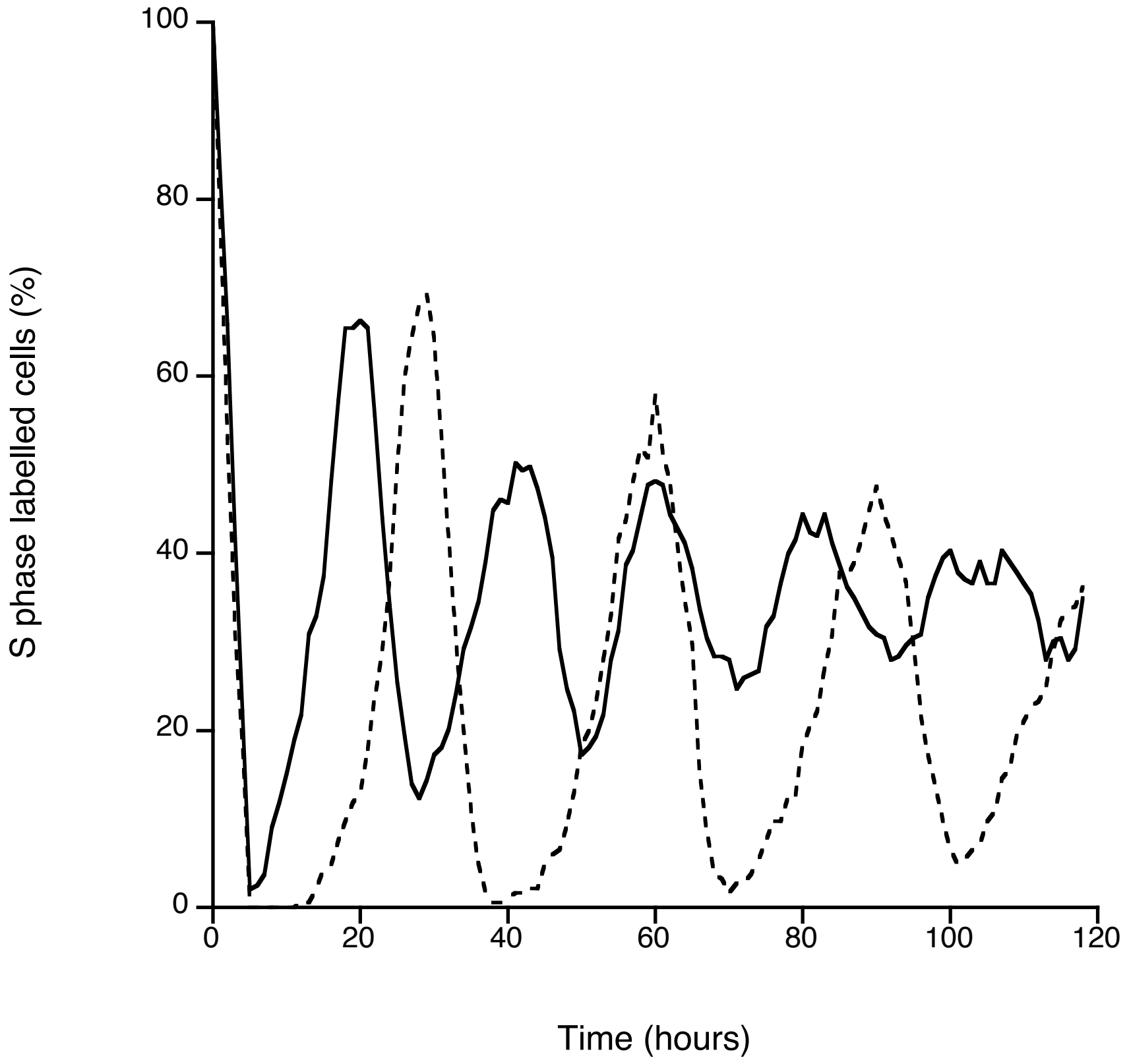

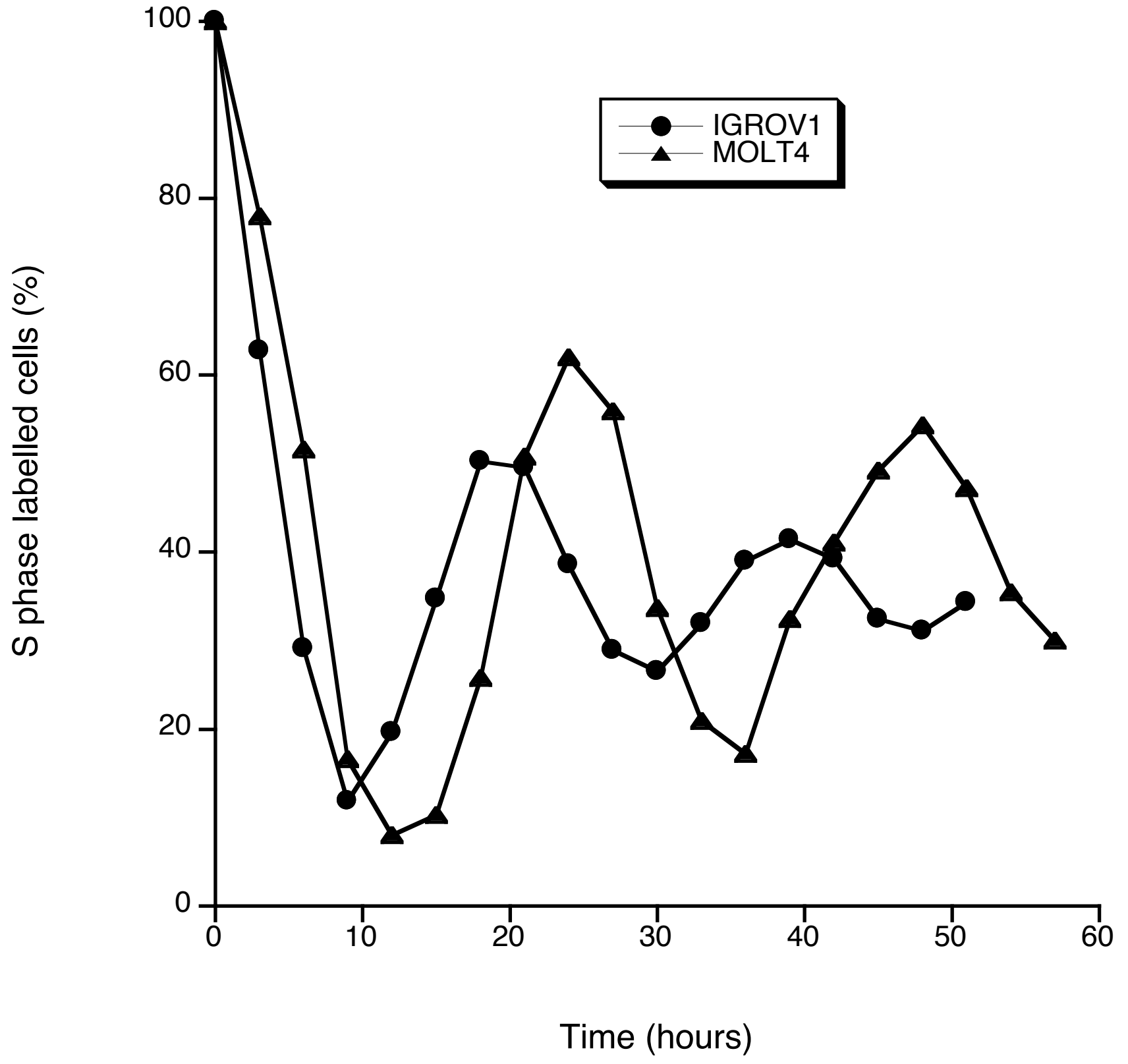

Fig.7 R.Chignola and E.Milotti